%% file: DSL.tex
\documentclass{article}
\usepackage[preprint, nonatbib]{neurips_2026} %  for submission to arXiv
\usepackage{bm}% bold math
\usepackage{dcolumn}% Align table columns on decimal point
\usepackage{lineno}
\usepackage{graphicx}
\usepackage{subfigure}
\usepackage{epstopdf}
\usepackage{float}
\usepackage{threeparttable}
\usepackage{overpic}
\usepackage{subfigure}
\usepackage{float}
\usepackage{enumitem}
\usepackage[normalem]{ulem}
\usepackage{cmap} %Make PDF files searchable and copyable
\usepackage{comment}
\usepackage{mathtext}
\usepackage{mathrsfs}
\usepackage{multirow}
\usepackage{xcolor}%0824
\usepackage{markdown}
\usepackage[utf8]{inputenc} % allow utf-8 input
\usepackage[T1]{fontenc}    % use 8-bit T1 fonts
\usepackage{url}            % simple URL typesetting
\usepackage{booktabs}       % professional-quality tables
\usepackage{amsfonts}       % blackboard math symbols
\usepackage{nicefrac}       % compact symbols for 1/2, etc.
\usepackage{microtype}      % microtypography
\usepackage{xcolor}         % colors
\usepackage{caption}
\usepackage{placeins}
\usepackage{makecell}
\usepackage{verbatim}
\usepackage{hyperref}
\definecolor{link_color}{RGB}{0,0,139}
\hypersetup{bookmarksnumbered=true,colorlinks,urlcolor=link_color,linkcolor=blue,anchorcolor=link_color,citecolor=link_color}
\hypersetup{breaklinks=true}
\usepackage[%
  backend=biber,       % Use biber backend
  style=phys,         % APS&AIP style, default is AIP style
  articletitle=false,biblabel=brackets,
  chaptertitle=false,pageranges=false,
    maxnames=3, minnames=3,  
]{biblatex}
\DefineBibliographyStrings{english}{
  andothers = {\emph{et al\adddot}} % \emph (italics)
}
\DeclareBibliographyDriver{misc}{%
  \usebibmacro{author}%
  \setunit{\addcomma\space}%
  \iffieldundef{eprint}
    {}
    {\href{https://arxiv.org/abs/\thefield{eprint}}{, arXiv:\thefield{eprint} (\thefield{year})}}%
  \finentry
}
\DeclareBibliographyDriver{online}{%
  \usebibmacro{author}%
  \setunit{\addcomma\addspace}%
  \printfield{title}%
  \setunit{\addperiod\space}%
  \iffieldundef{howpublished}
    {}
    {\iffieldundef{url}
       {\printfield{howpublished}}
       {\href{\thefield{url}}{\printfield{howpublished}}}}%
  \setunit{\addspace}%
  \printfield{year}%
  \finentry
}

\DeclareBibliographyDriver{software}{%
  \usebibmacro{author}%
  \setunit{\addcomma\space}%
  \iffieldundef{url}
    {}
    {\href{\thefield{url}}{, \thefield{title} (\thefield{year})}}%
  \setunit{\addspace}%
  \finentry
}

\usepackage{listings}
\begin{document}
\include{def-com}
\title{
HepScript: A Dual-Use DSL for Human-AI Collaborative Data Analysis Workflows in High-Energy Physics
}

\author{%
Junkun Jiao$^1$\thanks{Equal contribution.}\quad Tong Liu$^2$\footnotemark[1]\quad Ke Li$^{2,3}$\thanks{Corresponding author.}\quad Weimin Song$^1$\footnotemark[2]\quad Yipu Liao$^{2,3}$\\     \textbf{Bolun Zhang$^2$\quad Beijiang Liu$^{2,3}$\quad Chang-Zheng Yuan$^{2,3}$\quad Yue Sun$^{2,3}$}\\
$^1$ Jilin University, Changchun, Jilin, China \\
$^2$ Institute of High Energy Physics, CAS, Beijing, China\\
$^3$ University of Chinese Academy of Sciences, Beijing, China \\
\texttt{jiaojk1118@mails.jlu.edu.cn, weiminsong@jlu.edu.cn} \\
\{\texttt{like}, \texttt{liutong2016}, \texttt{liaoyp}, \texttt{zhangbolun}, \texttt{liubj}, \texttt{yuancz}, \texttt{yuesun}\}@\texttt{ihep.ac.cn}
}

%\date{\today}
\maketitle
	\begin{abstract}
         The escalating data scale in High-Energy Physics (HEP) fuels a growing aspiration for higher analytical efficiency. While Large Language Models (LLMs) offer a path toward automation via agentic AI, they struggle with complex scientific workflows that require deep domain knowledge and are tightly coupled to experiment-specific codebases. To address this, we introduce a methodology centered on HepScript, a dual-use Domain-Specific Language (DSL) for HEP data analysis workflows. HepScript serves as a shared formal interface, abstracting HEP analysis logic into a constrained syntax that is both intuitive for human experts and reliably generable by AI agents. First developed for the Beijing Spectrometer III (BESIII) experiment, HepScript hides the complexity of the underlying software stack, translating high-level analysis intent into low-level, production-ready code. In our case studies, this abstraction reduces the required human-written code by 93\%. Crucially, HepScript's constrained grammar defines a tractable action space, enabling AI agents to autonomously generate executable specifications for core analysis stages directly from published literature with a 95\% success rate. Our work demonstrates a scalable pathway toward human-AI collaborative systems, where a formally specified DSL acts as an unambiguous translation layer between human expertise, AI automation, and production environment, rendering previously intractable automation problems solvable.
        \end{abstract}

	%%%%%%%%%%%%%%%%%%%%%%%%%%%%%%%%%%%%%%%
	% 1. Introduction
	%%%%%%%%%%%%%%%%%%%%%%%%%%%%%%%%%%%%%%%
   \section{Introduction}\label{sec:introduction}

Scientific discovery is undergoing a profound paradigm shift toward an AI-driven fifth paradigm~\cite{Wang_2023}, where AI is expected to transform from a mere tool to an autonomous research collaborator~\cite{Ioannidis_5th_paradigm}. Recent advances in Large Language Models (LLMs)~\cite{2024ATSAS..1053070M} make this prospect tangible for data-intensive disciplines like High-Energy Physics (HEP), where the petabytes to exabytes of raw data accumulated annually~\cite{Weltman:2018zrl,ZurbanoFernandez:2020cco} fuel a growing aspiration for higher analytical efficiency, making AI automation a promising and perhaps necessary strategy. However, for LLM-powered agents to be practically useful, they must move beyond simple task automation to comprehending, planning, and executing complete multi-step scientific workflows. This remains a fundamental challenge, as LLMs often struggle with long-horizon planning~\cite{2024arXiv240201817K}, lack the deep, tacit domain knowledge required to structure these workflows, and cannot reliably interact with production environments without strictly predefined interfaces.

 HEP data analysis provides an ideal yet demanding testbed for autonomous AI capabilities. The workflows are intricate---spanning data collection, selection, visualization, interpretation, and statistical inference---and are tightly coupled to complex, experiment-specific software frameworks (e.g., BASF2~\cite{Kuhr:2018lps}). Because nearly all workflow procedures can be expressed as coding tasks, they are naturally well-suited for LLMs. However, our early exploration with the Beijing Spectrometer III (BESIII) experiment~\cite{ABLIKIM2010345} identified three critical barriers to autonomous HEP analysis: \textbf{(1) Formalizing Domain Knowledge:} How can the extensive, often implicit rules of HEP analysis be embedded into a standard format usable by AI agents? \textbf{(2) Integrating with Production Environment:} How can we bridge the semantic gap between a high-level analysis goal (e.g., reconstruct a particle) and the low-level, framework-specific code required to execute it? \textbf{(3) Constraining LLM Generation:} How can we guide LLMs to generate logically coherent, multi-step workflows that are guaranteed to compile and execute in a production environment?

\begin{figure*}[tb]
    \centering
    \begin{minipage}[t]{0.95\linewidth}
    \includegraphics[width=1.0\linewidth]{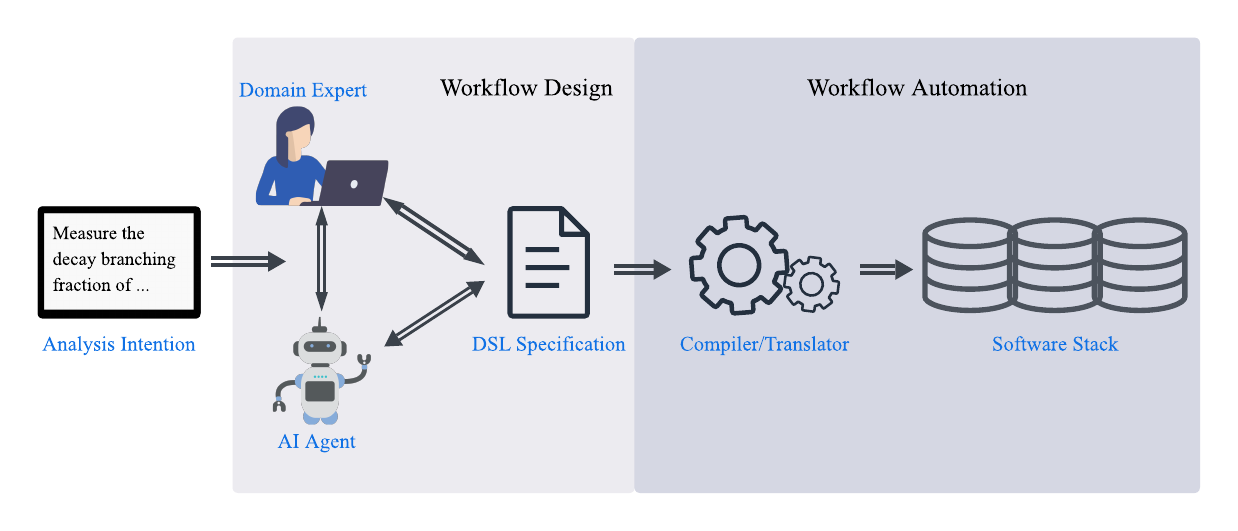}   
    \end{minipage}
    \caption{Dual-use DSL as a grounding mechanism. It provides an intuitive interface for human experts and a constrained action space for LLM agents. The formal grammar transforms open-ended code generation into a tractable sequence prediction task. A DSL processor (compiler/translator) then bridges the specification to production software stacks.}
    \label{fig:dsl_mechanism}
\end{figure*}

To overcome these barriers, we propose a grounding methodology utilizing a Domain-Specific Language (DSL), as illustrated in Fig.~\ref{fig:dsl_mechanism}. Our key insight is that a well-designed DSL can collapse the semantic gap between analysis intent and machine execution. By abstracting HEP data analysis logic into a constrained, high-level syntax, the DSL serves a dual purpose: it provides an intuitive interface for physicists, and it defines a tractable, bounded action space for an LLM agent. The DSL's formal grammar acts as a guardrail, transforming an open-ended code generation problem into a constrained sequence prediction task that is far more amenable to few-shot learning. This approach simultaneously solves all three barriers: formalizing domain knowledge into the DSL's grammar, integrating with the software framework via a DSL processor, and providing necessary, verifiable structures for LLM generation. Through this, it builds a formal model for HEP data analysis, shifting the automation bottleneck from \textit{how to conduct} an analysis to \textit{what to specify}.

\subsection{Contributions}
In this paper, we introduce and validate a methodology for grounding autonomous agents in complex scientific workflows using a dual-use DSL. Our contributions are fourfold:
\begin{enumerate}
    \item \textbf{Core Insight:} We demonstrate that a well-designed DSL can collapse the unbounded action space of framework-specific code, transforming open-ended LLM generation into a highly constrained, reliable sequence prediction task.
    \item \textbf{Concrete Instantiation:} We present HepScript, a Ruby-embedded DSL co-designed with LLM assistance for the BESIII experiment, which abstracts the complex underlying software system into a formal, executable specification language.
    \item \textbf{Empirical Validation:} Through systematic evaluations, we show that HepScript reduces human coding effort by 93\% and enables LLMs to autonomously generate correct, executable specifications for core analysis stages with a 95\% success rate after agentic retries. 
    \item \textbf{Generalizable Methodology:} We provide a documented methodology---covering DSL design, implementation details, generation strategies, evaluation setups, and a proposed self-evolutionary mechanism---that can be adapted to other scientific domains where framework-heavy workflows currently preclude automation.
\end{enumerate}

\subsection{Related Works}
Recent studies~\cite{2025arXiv251207785G,2025arXiv251215867M,2026arXiv260305735B,2026arXiv260320179M} have explored automating HEP workflows. Ref.~\cite{2025arXiv251207785G} combines a relatively fixed workflow manager with a code-generating agent. Ref.~\cite{2025arXiv251215867M} formalizes individual workflow phases as schema-validated tools that consume run cards and output JSONL data, with agent orchestration in a human-supervised loop. Ref.~\cite{2026arXiv260305735B} employs an LLM as a ``graduate student" supervised by human physicists to execute a fixed analysis, lacking a generalizable agentic framework. Building on that direction, Ref.~\cite{2026arXiv260320179M} introduces more domain-specific rules and tools but delegates the oversight tasks to specialized AI agents. While these prior efforts offer promising architectures, none defines a formal, unified representation of a complete workflow that is simultaneously human-friendly, AI-generable, and software-integrated. Our work on HepScript explores whether a DSL could fill this role. HepScript is not a replacement for these agentic systems; rather, it can be integrated into them as a shared action space, an intermediate language, or a validation layer. 

In the HEP community, early DSLs targeted specific functionalities such as database querying or simple data processing. To our knowledge, only the recently emerged ADL~\cite{2022arXiv220309886P}, b2luigi~\cite{alexander_heidelbach_2025_15229241}, and FLARE~\cite{CooperHarris:2025lqd} have the potential to capture complete analysis pipelines. However, b2luigi and FLARE are Python libraries focused primarily on workflow orchestration, while ADL is designed to decouple physics logic from framework technicalities to facilitate accessibility, analysis (re)interpretation, and preservation. Critically, these tools are not intermediate representations explicitly built for human-agent collaboration. HepScript is a preliminary attempt to explore this direction within the BESIII context and can be encapsulated as a tool or ``skill"~\cite{Anthropic2025} for multi-agent systems.
   
	%%%%%%%%%%%%%%%%%%%%%%%%%%%%%%%%%%%%%%%%
	% 2. DSL Design and Features
	%%%%%%%%%%%%%%%%%%%%%%%%%%%%%%%%%%%%%%%%

\section{Design and Evolution of HepScript}\label{sec:dsl_design}
\subsection{Code Generation Architecture}
HepScript is designed around a code-generation architecture to balance minimal development effort against maximal abstraction utility. A typical BESIII data analysis workflow involves two distinct software suites: the BESIII Offline Software System (BOSS)~\cite{Zou:2024pmc} for simulation, reconstruction, and basic data selection, and ROOT~\cite{Antcheva:2009zz} for subsequent high-level data selection, statistical inference, and visualization. The inherent complexity of BOSS and the flexibility of ROOT make creating a single compiled or interpreted language spanning both software suites impractical. Therefore, we adopt a code generation architecture. HepScript does not execute the workflow directly. Instead, it serves as a high-level specification language. A dedicated processor reads the HepScript file and generates the required analysis code snippets for different workflow procedures. These snippets are then orchestrated and executed---either manually or by an automated agent---in the correct sequence. This approach decouples HepScript from the runtime environments of BOSS and ROOT, enabling code generation in multiple target languages (such as C++ for BOSS algorithms and ROOT scripts, Bash for job configuration, or Python for scientific libraries) depending on the analysis phase. Furthermore, the intermediate DSL processor facilitates autonomous error recovery by providing detailed diagnostic feedback and correction suggestions during agentic retries. Finally, this design ensures full traceability for human analysts. Experts can inspect and modify the generated code, trace problems back to the original HepScript specification, and intervene as necessary. This dual transparency is essential for building reliable human-AI collaborative systems, acting as a safeguard against the probabilistic nature of LLM-generated components.

  Given this architecture, we implemented HepScript as an embedded DSL within a general-purpose host language. This leverages the host's existing syntax and parser, accelerating development compared to creating an external DSL with dedicated parsing tools (e.g., Yacc/Bison) and facilitating LLM generation. To ensure dual-use capability, HepScript adheres to three core constraints:
  \textbf{(C1) Readability:} The syntax must be declarative and intuitive, shielding users from framework complexities to ensure human adoption and reliable LLM generation. 
  \textbf{(C2) Modularity:} HepScript must provide discrete, reusable blocks corresponding to BESIII analysis stages, enabling workflow composition and task decomposition.
  \textbf{(C3) Host Language Suitability:} The host language must support syntactic flexibility and meta-programming for an elegant DSL capable of complex code generation.
  
After evaluating potential hosts (e.g., Python, Ruby, Lisp, Smalltalk) against these constraints, we selected Ruby for its exceptionally readable and flexible syntax, native support for fluent interfaces via method chaining, powerful meta-programming capabilities, and rich ecosystem.

\subsection{LLM-assisted HepScript Design}
To ensure HepScript addresses real-world analysis needs, we grounded its design in actual BESIII physics publications. We began with a curated corpus of 20 published BESIII papers describing analyses with simple and linear workflows. For each paper, we prompted an LLM with the full text and explicit instructions to: (1) extract the complete data analysis workflow, and (2) formalize it as structured, Ruby-embedded DSL pseudo-code that captures the analysis logic in a declarative style.
The initial DSL outputs were manually analyzed to identify frequent semantic patterns (e.g., ``select photon", ``particle identification"). This informed an iterative, human-in-the-loop refinement process. We continuously reviewed the LLM's proposed syntax, generalized recurring patterns into core DSL constructs, and refined the prompt based on observed shortcomings. This cycle converged on an optimized DSL grammar that provides an intuitive, fluent interface for expressing analysis logic.

The final HepScript's structure is synthesized to adhere to the five stages of a typical BESIII analysis:
\begin{enumerate}
    \item \textbf{Dataset Preparation:} Declaring the real data and simulated Monte Carlo (MC) samples.
    \item \textbf{Base Selection:} Applying basic data selection criteria to separate candidate signal events from background events.
     \item \textbf{Advanced Selection:} Purifying the candidate signal sample via optimized, analysis-specific selection criteria to suppress background pollution.
   \item \textbf{Visualization:} Presenting and interpreting the data, such as generating comparative figures of invariant mass distributions between simulated samples and real data.
    \item \textbf{Statistical Analysis:} Perform statistical analyses, such as fitting distributions and computing significances for physical measurements.
\end{enumerate}
The first two stages generate code for the BOSS, while the latter three target ROOT. An example of the resulting HepScript's grammar is provided in the Appendix~\ref{sec:appendix_A}. Through this process, we have successfully distilled the domain model of BESIII data analysis and instantiated it into the HepScript representation.

   %%%%%%%%%%%%%%%%%%%%%%%%%%%%%%%%%%%%%%%%
   % 3. DSL implementation
   %%%%%%%%%%%%%%%%%%%%%%%%%%%%%%%%%%%%%%%%   
   \section{\label{Sec:dsl_implementation}Implementation}
\begin{figure*}[tb]
    \centering
    \begin{minipage}[t]{0.99\linewidth}
    \includegraphics[width=1.0\linewidth]{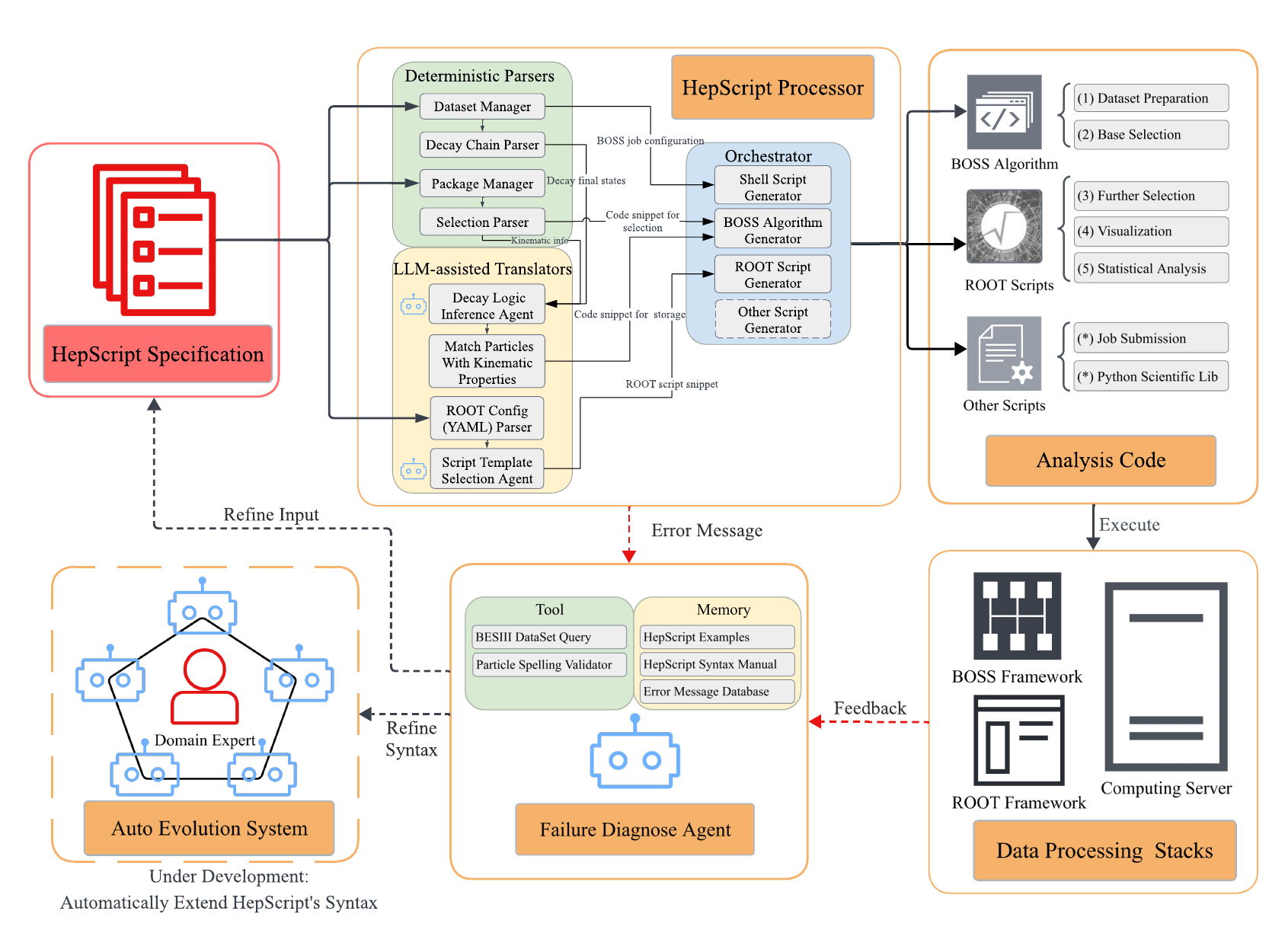}        
    \end{minipage}
    \caption{Overview of the HepScript-grounded data analysis workflow for BESIII. The HepScript specification serves as a unified interface, specifying all instructions required to complete the analysis. A dedicated processor generates and orchestrates code snippets for multiple target systems: BOSS (dataset preparation and base selection), ROOT (advanced selection, visualization, and statistical analysis), and auxiliary scripts (e.g., Linux Shell, Python). The generated code is executed within corresponding data processing stacks. Upon failure, the outputs (processor errors and system logs) are fed to a diagnostic agent, which either refines the HepScript specification or invokes the auto-evolution system to extend HepScript's syntax. Human experts can supervise and intervene at any stage, enabling rapid prototyping and debugging within a human-AI collaborative system.}
    \label{fig:dsl_workflow}
\end{figure*}

To realize the code-generation architecture described in Sec.~\ref{sec:dsl_design}, we built a DSL processor that translates HepScript specifications into target analysis code. The processor adopts a hybrid generation strategy with three approaches, selected based on the nature of the target code:
\begin{itemize}
    \item \textbf{Templated Generation:} For target code with extensive static structure and few dynamic variables (e.g., selecting charged tracks in base selection), we use template files containing placeholder keys (e.g., \textit{\{\{key\}\}}). The processor substitutes these placeholders with analysis-specific values extracted from the HepScript specification. This method supports iterative substitution, where the substituted value may contain other placeholders, enabling the handling of complex, nested code structures.
    
\item \textbf{Translator-Based Generation:} For target code with highly flexible or semantically complex syntax that is ill-suited to rigid templates, we implement specialized Ruby translator classes. These parsers process the HepScript content and apply custom logic to generate the corresponding target code, offering precise structural control.

\item \textbf{LLM-Assisted Generation:} We employ LLMs in two scenarios, where translation logic is dynamic and deeply analysis-dependent:
    \begin{enumerate}
        \item \textbf{Cascade Decay Logic:} When storing physical variables (e.g., invariant masses) after base selection, the correct mapping between final-state particles and intermediate resonances can be ambiguous. For example, a decay chain containing multiple photons requires identifying which pairs originate from $\pi^0$ or $\eta$ meson decays. We prompt an LLM to determine the optimal particle combination based on expert-provided logic (e.g., minimizing $(M(\gamma_{i}\gamma_{j})-m_{\pi^{0}})^{2}+(M(\gamma_{k}\gamma_{l})-m_{\eta})^{2}$, where the subscripts $i$, $j$, $k$, and $l$ denote different photons, $M(\gamma\gamma)$ is the invariant mass of two-photon combination, and $m_{\pi^{0}}(m_{\eta})$ is the nominal mass of $\pi^{0}(\eta)$). The LLM outputs well-commented source code for variable storage (see Appendix~\ref{sec:appendix_B}).
        \item \textbf{ROOT Script Generation:} ROOT-based tasks are highly domain-specific. However, because ROOT is open-source, the underlying logic of individual tasks is relatively concise. We instruct LLMs to generate the required ROOT scripts from formalized YAML inputs and curated exemplars.
    \end{enumerate}
    
\end{itemize}
These three approaches are integrated within a syntax-directed translation framework. HepScript grammar functions map directly to corresponding Ruby classes and methods within the processor. The final system allows a HepScript specification to be executed directly from the command line, automatically generating the complete suite of required analysis code for BOSS and ROOT. The resulting workflow is illustrated in Fig.~\ref{fig:dsl_workflow}.
   
	%%%%%%%%%%%%%%%%%%%%%%%%%%%%%%%%%%%%%%%%
	% 4. DSL generation
	%%%%%%%%%%%%%%%%%%%%%%%%%%%%%%%%%%%%%%%%
\section{Generating HepScript: In-context Learning}\label{Sec:dsl_generation}
A core claim of this work is that HepScript's constrained syntax makes it generable by LLMs. However, generating any DSL requires effective strategies for guiding the LLM, especially when embedding specialized domain knowledge. While fine-tuning (FT) is a viable option, it is expensive, difficult to optimize, and hard to adapt to a rapidly evolving DSL. Furthermore, recent studies reveal that optimized Retrieval-Augmented Generation (RAG) can achieve comparable or better syntactic correctness than FT~\cite{2024arXiv240702742B}. Therefore, we initially focused on in-context learning via RAG. However, we found that the standard RAG paradigm, which retrieves examples based on the semantic similarity of natural language queries, is insufficient for our domain.

\subsection{Limits of Semantic Similarity Retrieval for HEP Workflows}

Standard RAG systems embed a user's query and retrieve the most semantically similar examples from a database. While effective for many tasks, HEP analysis is exquisitely sensitive to the details of the selection criteria and analysis-dependent statistical methods. An analysis involving two photons plus a low-energy transition photon in the final states is physically distinct from one involving only two photons, even if their natural language descriptions are close. Using retrieved examples based purely on linguistic similarity leads to incorrect selection criteria, flawed particle combinations, or invalid fit strategies, corrupting the measurement.

The core problem is that semantic similarity in natural language does not align with the formal structural similarity that governs physical correctness. The "distance" between two analyses is defined by the topology of the particle decay chain (e.g., the number and types of final-state particles, intermediate resonances), not the words used to describe them. Developing a structure-aware retriever for physical isomorphisms remains an open and challenging problem.

\subsection{A Baseline: Comprehensive Single-Shot Grounding}
In the absence of a structure-aware retriever, we established a strong baseline by providing the LLM with exhaustive grounding information in a single, long-context prompt (typically 30-40k tokens). This approach, while not scalable indefinitely, demonstrates the feasibility of the approach within the current context window and provides an upper bound on performance when all necessary syntax information is available. Our grounding package consists of two components: \textbf{A Comprehensive Workflow Example:} We provide a complete HepScript specification for a canonical BESIII analysis, demonstrating the intended use of all core language constructs in an end-to-end workflow. \textbf{A Complete Grammar Reference (in YARD):} We include YARD-formatted documentation for the public HepScript APIs (see Appendix~\ref{sec:appendix_C}). This serves as a formal grammar specification, detailing every function, its parameters, return values, and usage examples.

The LLM is instructed to study the comprehensive example and consult the YARD documentation as an authoritative reference during generation. This transforms the generation task from open-ended synthesis to a constrained, reference-driven translation task. As our evaluation (Sec.~\ref{Sec:dsl_evaluation}) shows, a well-designed DSL and comprehensive documentation enable reliable code generation even with a single-shot, ultra-long prompt. The full prompt is shown in Appendix~\ref{sec:appendix_D}.
	%%%%%%%%%%%%%%%%%%%%%%%%%%%%%%%%%%%%%%%%
	% 5. DSL evaluation
	%%%%%%%%%%%%%%%%%%%%%%%%%%%%%%%%%%%%%%%%
\section{Evaluation}\label{Sec:dsl_evaluation}
We evaluate HepScript along two dimensions aligned with its dual-use design: (1) as a human-facing abstraction, assessed by its ability to faithfully represent workflows and reduce coding effort; and (2) as an AI-facing interface, assessed by how well LLMs can generate correct HepScript specifications from domain literature.

\subsection{Experimental Setup}\label{Sec:dsl_evaluation_setup}
We selected the first fifty BESIII papers from arXiv (from 2009 to the evaluation date), sorted by publication date. Each paper was categorized by research methodology using DeepSeek-V3~\cite{deepseekai2024deepseekv3technicalreport}; we excluded papers relying on methods beyond HepScript's current expressiveness (e.g., deep learning). Such papers constitute a small fraction of the corpus. After filtering, 45 papers remained. For each, one domain expert wrote the corresponding HepScript specification, and a second expert verified it to establish our ground truth.

We define three evaluation metrics for HepScript specifications: (i) \textbf{Syntax correctness:} The HepScript processor accepts the specifications without errors; (ii) \textbf{Logical correctness:} the described workflow is physically meaningful and adheres to the intended analysis, as verified by an expert; (iii) \textbf{Compilation (execution) success:} the generated code compiles (runs) successfully in its respective environment, producing expected outputs. A specification is considered overall successful only if it satisfies all three metrics.

The evaluation uses five LLMs: DeepSeek-V3, DeepSeek-R1~\cite{Guo_2025}, GPT-4o~\cite{openai2024gpt4ocard}, GLM-4.7~\cite{5team2025glm45agenticreasoningcoding}, and Qwen3-Max~\cite{yang2025qwen3technicalreport}. Among these, DeepSeek-R1 and GLM-4.7 are advanced reasoning models.

\subsection{Evaluating HepScript Processor}
We split this evaluation into three parts: BOSS code generation, ROOT code generation, and HepScript generation within an agentic loop.

\textbf{BOSS Code Generation:}
The processor's core translation engine (templates + translators) is deterministic. To verify its correctness, we ran all human-written HepScript specifications through the processor with the LLM-assisted component (see Sec.~\ref{Sec:dsl_implementation}) disabled, yielding 63 algorithm packages (a HepScript specification may produce multiple packages). In all cases, the generated BOSS code compiled without errors, confirming that the processor's core logic is sound.

\textbf{ROOT Code Generation:}
Because preparing test-ready datasets for BESIII requires substantial computational resources, we limited the ROOT scripts execution evaluation to two representative analyses: $J/\psi \to \gamma \pi^{+}\pi^{-}\eta'$ and $\psi'\to \pi^{+}\pi^{-}J/\psi(\to \gamma p \bar{p})$. For each, we wrote a HepScript specification and executed the full analysis pipeline, from dataset preparation to reproducing the original paper's figures.

\textbf{HepScript Generation (Agentic Loop):}
Given the full text of a paper, we used reasoning models to generate HepScript specifications following the method in Sec.~\ref{Sec:dsl_generation}. To evaluate this capability at scale, we used the 45-paper corpus, focusing on dataset preparation and base-selection stages (most code-intensive portions). Executing these specifications yielded 72 algorithm packages.

\subsection{Evaluation Results}

\textbf{BOSS Code Generation:}
 For the variable storage sub-task, we enabled the LLM-assisted component to compare different LLMs. Table~\ref{tab:evaluation_result} reports the success rates and standard errors, assuming binomial distribution with a uniform prior following a Bayesian inference approach~\cite{2009A}. All models achieved high initial success rates except DeepSeek-R1 (87.7\%), whose failures stemmed from four timeout errors due to excessive "overthinking" and three variable name mismatches (common failure across LLMs). Critically, after one agentic retry with error feedback, all models achieved a near-perfect success rate (98.5\%), demonstrating the effectiveness of our LLM-assisted translation strategy. 

\begin{table}[tb]
\centering
\caption{Comparison of success rates (SR) for human-written versus LLM-generated HepScript specifications (BOSS code only). For each LLM, $N_{S}$ is the number of successful packages and $N_{F}$ is the number of failed packages. “One Retry SR” and “Three Retry SR” denote success rates after one and three agentic iterations with error message feedback, respectively.}
\vspace{1\baselineskip} 
\label{tab:evaluation_result}
\resizebox{\textwidth}{!}{%
\begin{tabular}{ccccccc}
\toprule
HepScript&{LLM} & {$N_{S}$} & {$N_{F}$} & {SR (\%)} & {One Retry SR (\%)} &{Three Retries SR (\%)}\\
\midrule
\multirow{4}{*}{\makecell{Human-written\\(63 Packages)}}&Deepseek-V3 & 62 & 1 & $96.9\pm2.1$ & $98.5\pm{1.5}$&- \\
&Deepseek-R1 & 56 & 7  & $87.7\pm4.0$ & $98.5\pm{1.5}$&- \\
&gpt-4o & 62 & 1 & $96.9\pm2.1$& $98.5\pm{1.5}$&- \\
&GLM-4.7 & 62 & 1 & $96.9\pm2.1$ & $98.5\pm{1.5}$&- \\
&Qwen3-Max & 60 & 3 & $93.8\pm3.0$& $98.5\pm{1.5}$&- \\
\midrule
%\cmidrule{r}{1-7}
\multirow{2}{*}{\makecell{LLM-generated\\(72 Packages)}}  & Deepseek-R1 & 34 & 38 & $47.3\pm5.8$ & $87.8\pm3.8$&$94.6\pm2.6$ \\ 
&GLM-4.7 & 31 & 41 & $43.2\pm5.7$ & $90.5\pm3.4$&$95.9\pm2.3$\\ 
\bottomrule
\end{tabular}
}
\end{table}

\textbf{HepScript Generation:}
 As shown in Table~\ref{tab:evaluation_result}, 47.3\% (43.2\%) of specifications generated by Deepseek-R1 (GLM-4.7) succeeded on the first attempt. Failures primarily arise from (1) incorrect usage of HepScript's syntax (76\%) such as calling undefined methods or passing an incorrect number of arguments, and (2) physics misinterpretations (24\%), such as omitting a kinematic fit that constrains the final state to the center-of-mass energy, or referring to particles that have not yet been reconstructed. After a single agentic retry with error feedback, the success rates jumped to 87.8\% (90.5\%). After three iterations, the success rates are about 95\% for both LLMs. The remaining failures involved multiple errors that require many more retries. These results prove that a well-designed DSL, paired with an iterative agentic loop, enables highly reliable automation---a promising outcome for human-AI collaborative systems. 
 
\begin{figure*}[tb]
\centering
\begin{minipage}[t]{0.99\linewidth}
\includegraphics[width=0.47\linewidth]{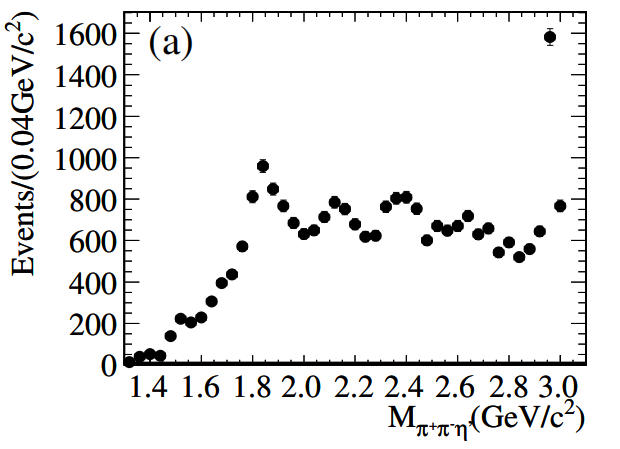}
\includegraphics[width=0.47\linewidth]{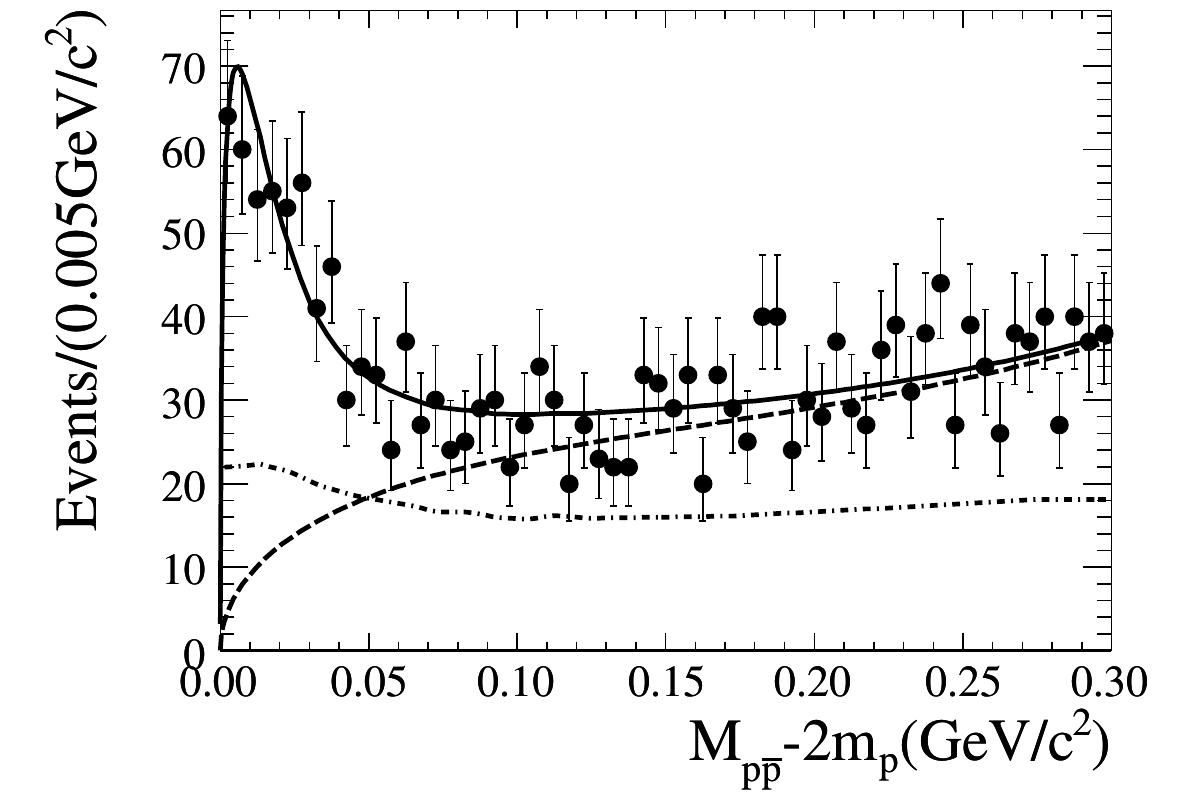}
\end{minipage}
\begin{minipage}[t]{0.99\linewidth}
\includegraphics[width=0.47\linewidth]{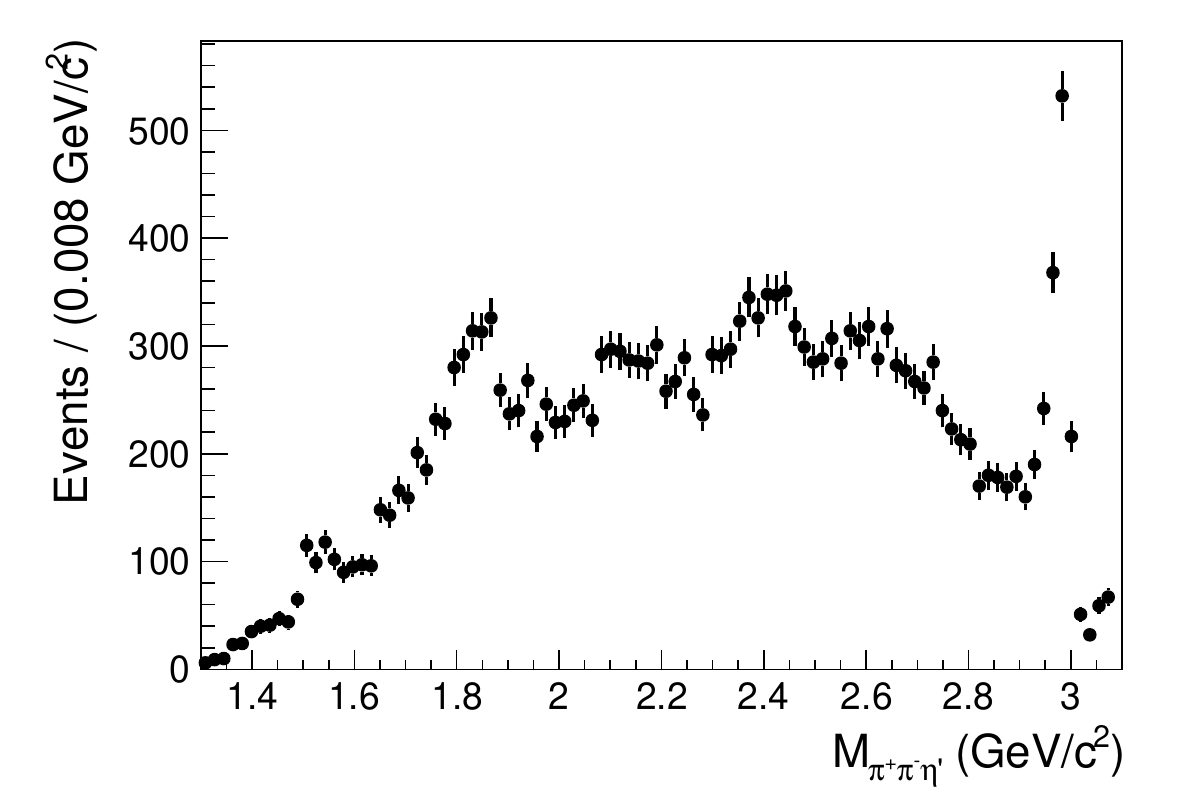}
\includegraphics[width=0.47\linewidth]{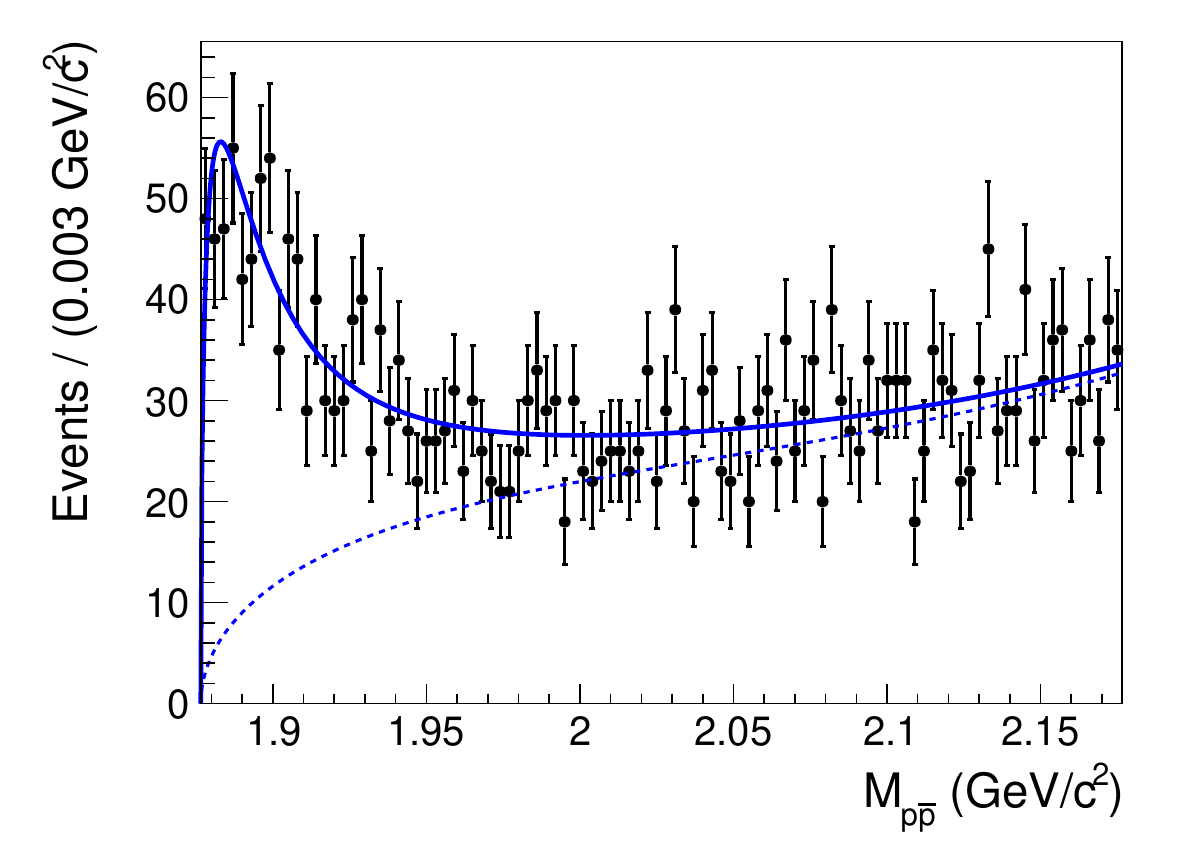}
\end{minipage}

\caption{Comparison of original and reproduced figures. Top panels: original figures from BESIII publications: (left) $J/\psi \to \gamma \pi^{+}\pi^{-}\eta'$~\cite{BESIII:2010gmv} and (right) $\psi'\to \pi^{+}\pi^{-}J/\psi(\to \gamma p \bar{p})$~\cite{BESIII:2010vwa}. Bottom panels: reproduced figures. In the right column, the dashed and solid curves denote the fitted background
function and the fit result, respectively. Note that these reproduced figures serve strictly to validate the automated pipeline and should not be interpreted as official physics results.} %In the right column, discrepancies between original and reproduced distributions arise from the MC simulation of the physical process.
\label{fig:root_evaluation}
\end{figure*}

\textbf{ROOT Code Generation:}
In both case studies, the generated ROOT scripts execute without error and reproduce figures matching the original papers, as shown in Fig.~\ref{fig:root_evaluation}. While the sample size is limited, the successful execution suggests that LLMs can effectively translate HepScript tasks into analysis-ready ROOT code. We emphasize that these reproduced figures serve strictly as a technical validation of the automated pipeline and should not be interpreted as official physics results.

\textbf{Reduction in Coding Effort:}
Across the two case studies, HepScript reduced the volume of analysis code (measured by character count, excluding comments and blank lines) written by humans by an average of 93\%. %(range: 92\%-94\%).
This reduction comes primarily from eliminating boilerplate BOSS code and repetitive ROOT plotting routines. For human experts, this translates to faster prototyping and fewer low-level errors; for AI agents, it defines a dramatically smaller action space.

\subsection{Limitations of the Evaluation}
\label{sec:limitation of the evaluation}

\textbf{Benchmark Scope.} Our corpus of 45 papers, while representative of typical BESIII measurements, does not cover the full diversity of HEP workflows. Notably absent are complex analyses such as cross-section measurements and amplitude analyses, which may require additional language constructs. This reveals a critical need for the community: a structured benchmark for HEP analysis workflows.

\textbf{Analysis Logic Evaluation.} Assessing the logical correctness of LLM-generated HepScript specifications requires an expert review process, which is neither scalable nor objective. However, this manual evaluation remains unavoidable, as it demands deep HEP knowledge that cannot yet be automated.

\textbf{Computational Cost.} Full execution-based evaluation, particularly for ROOT scripts, requires preparing large datasets and running computationally intensive simulations. This limits our ROOT evaluation to two case studies. Developing lightweight surrogate metrics that correlate with execution correctness, or creating small-scale "toy" datasets that preserve the essential structure of analyses while reducing computational demands, would enable broader validation.

	%%%%%%%%%%%%%%%%%%%%%%%%%%%%%%%%%%%%%%%%
	% 6. Discussion and Future works
	%%%%%%%%%%%%%%%%%%%%%%%%%%%%%%%%%%%%%%%
   \section{Discussion and Future Work}\label{sec:discussion}
   
\subsection{Expressiveness vs. Abstraction}
A fundamental challenge in designing a dual-use DSL is balancing expressiveness against LLM-generability. Higher abstraction simplifies LLM generation, but necessarily reduces fine-grained control over the workflow; lower abstraction preserves flexibility but increases the risk of LLM hallucination. Finding the optimal balance requires iterative refinements guided by real-world analysis requirements. The current HepScript covers a subset of BESIII analysis types but lacks constructs for systematic uncertainty estimation, a critical component of any physics result. Future work will expand HepScript's expressiveness to support more sophisticated analyses including systematic uncertainty estimation. We will further explore a self-evolutionary mechanism that autonomously extends HepScript's grammar. The mechanism would ingest domain papers and identify coverage gaps from practical usage, as depicted in Fig.~\ref{fig:dsl_workflow}. Initially, human experts would validate and integrate the proposed syntax extensions, with the system moving toward full automation as it matures.

\subsection{Toward Structure-Aware Retrieval for LLM Generation}
Our current approach of HepScript generation (a comprehensive example and YARD syntax reference) is not a scalable long-term solution as HepScript grows. This work therefore identifies a clear research agenda: the development of structure-aware retrieval mechanisms that operate on the formal topology of particle physics processes rather than semantic text similarity. We believe that developing structure-aware embeddings or domain-specific retrieval algorithms for scientific workflows is a crucial next step for any multi-agent system for scientific discovery, and is our top priority.

\subsection{Development of an Agentic Memory Mechanism}
Each HepScript specification formalizes a complete BESIII analysis pipeline, collectively forming a highly structured knowledge database. Viewed through the lens of harness engineering, this repository extends beyond a static reference to become a dynamic memory module for future multi-agent systems~\cite{Li:2026krn}. We envision a self-reinforcing cognitive loop where memory, skills, and protocols continuously interact~\cite{2026arXiv260408224Z}. Specifically, the database would supply structural evidence and successful execution trajectories, allowing agents to distill stored experience into reusable analytical procedures (skills) for new workflows. These skills would then transition from abstract reasoning into governed action via HepScript's constrained grammar and the processor's validation loop, which act as strict protocols providing typed interfaces and boundary checks that ensure safe, verified execution. Finally, closing the loop through result assimilation, successfully executed analyses would be normalized by the protocol layer and written back into memory as new, unified specifications. As this self-reinforcing cycle accelerates, the expanding memory repository would unlock frontier possibilities for machine learning on the workflows themselves---such as training models to predict effective selection criteria for specific particle final states, or to autonomously route protocol strategies based on historical success rates.
 
	%%%%%%%%%%%%%%%%%%%%%%%%%%%%%%%%%%%%%%%%
	% 7. Conclusion
	%%%%%%%%%%%%%%%%%%%%%%%%%%%%%%%%%%%%%%%%
	\section{Conclusion}\label{sec:conclusion}
 This paper demonstrates that DSL-grounded abstraction is a powerful strategy for automating intricate, framework-bound scientific workflows, using high-energy physics data analysis as a rigorous testbed. We show that a carefully designed DSL collapses the action space and acts as a shared representation layer that bridges human expertise, AI automation, and production environments. Ultimately, it shifts the fundamental problem of automation from how to execute an analysis to what to specify. We instantiated this approach with HepScript, a Ruby-embedded DSL for the BESIII experiment. The evaluation shows that HepScript reduces manual coding effort by 93\% and, more critically, enables AI agents to autonomously generate valid specifications from published literature with a 95\% success rate after agentic retries. These results validate the potential of a dual-use DSL in human-AI collaborative systems, where agents propose workflows and humans---or specialized diagnostic agents---provide corrective feedback.

Looking forward, this work lays the foundation for a self-evolutionary system that could autonomously abstract DSL grammar from domain literature, implement new language constructs, generate DSL specifications for real-world analyses, and iteratively refine the syntax based on deployment feedback. Such a system would close the loop, enabling the DSL and its associated agents to continuously improve alongside the domain. While developed and evaluated within the BESIII experiment, the principles of this methodology are readily extensible to other HEP experiments, such as Belle II~\cite{Belle-II:2018jsg} or LHCb~\cite{LHCb:2008vvz}, as well as other data-intensive scientific domains. We hope this work inspires such efforts.

 %%%%%%%%%%%%%%%%%%%%%%%%%%%%%%%%%%%%%%%%%%%%%%%%%%%%%%%%%%%
	% ACKNOWLEDGMENTS
	%%%%%%%%%%%%%%%%%%%%%%%%%%%%%%%%%%%%%%%%%%%%%%%%%%%%%%%%%%%

   \input{Acknowledgement.tex}
   
	%%%%%%%%%%%%%%%%%%%%%%%%%%%%%%%%%%%%%%%%%%%%%%%%%%%%%%%%%%%
	% Appendix
	%%%%%%%%%%%%%%%%%%%%%%%%%%%%%%%%%%%%%%%%%%%%%%%%%%%%%%%%%%%
\printbibliography

\newpage

\FloatBarrier
\newpage
\appendix
\section*{Technical appendices and supplementary material}
 \section{HepScript Example}\label{sec:appendix_A}
% DSL example (First part of the figure)
\begin{figure*}[h]
    \centering
    \begin{minipage}[t]{0.95\linewidth}
        \includegraphics[width=1.0\linewidth]{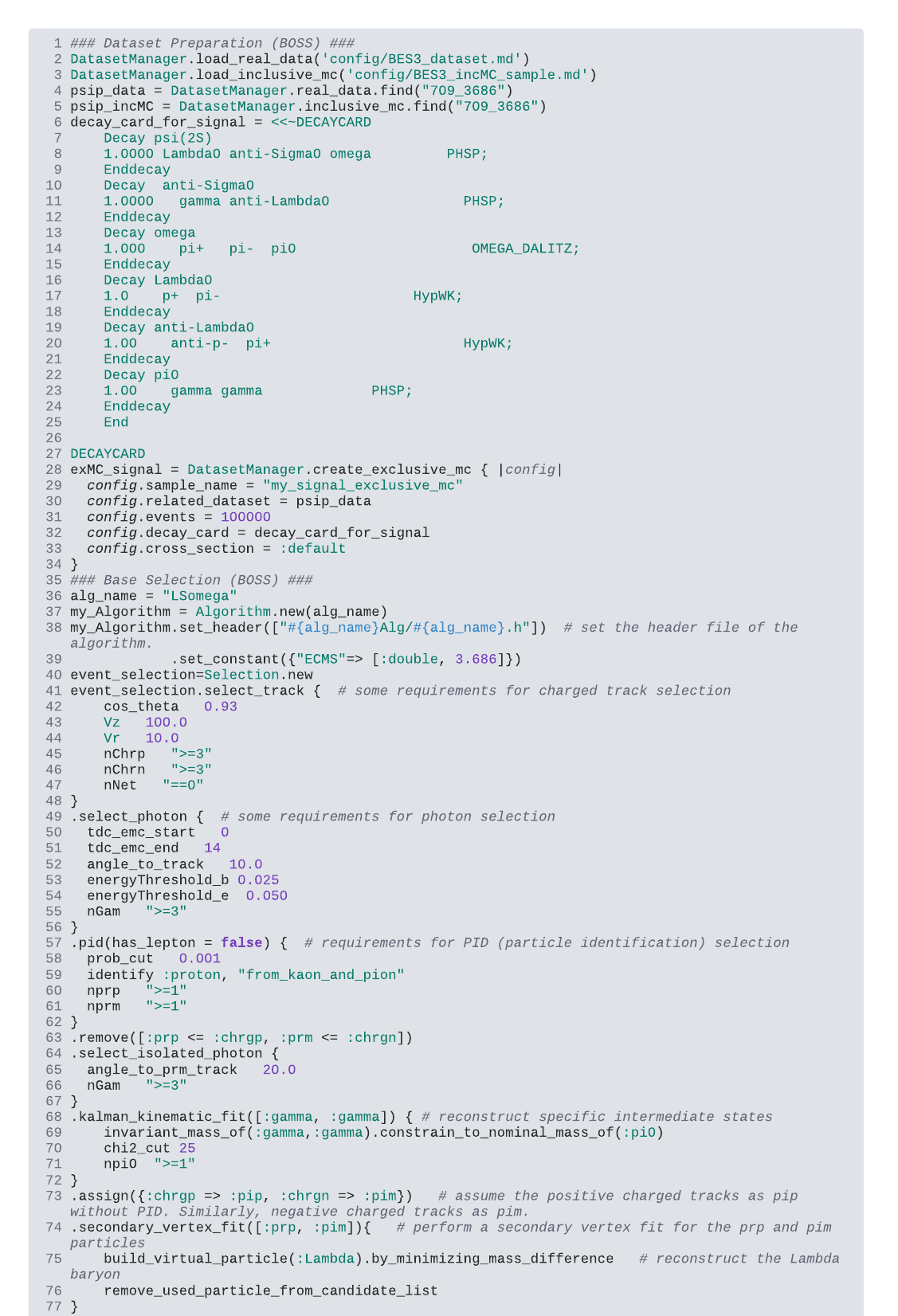}        
    \end{minipage}
    \caption{Example of HepScript's grammar. The involved decay process is $\psi(3686)\rightarrow \Lambda \bar{\Sigma}^{0}\omega$.}
    \label{fig:dsl_example}
\end{figure*}
\FloatBarrier
% Second part of the figure (continued on next page)
\begin{figure*}[tb]
    \ContinuedFloat %
    \centering
    \begin{minipage}[t]{0.95\linewidth}
        \includegraphics[width=1.0\linewidth]{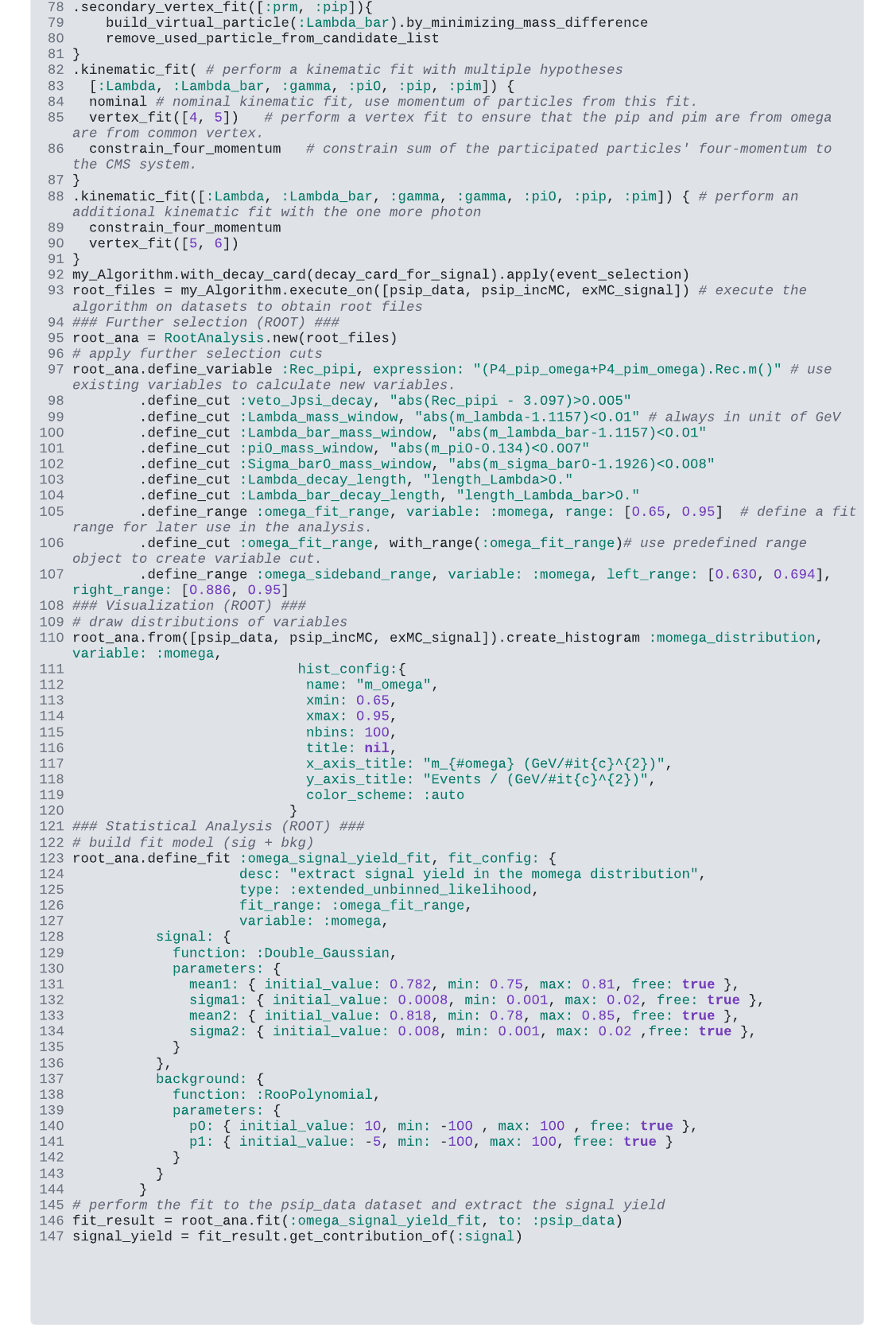}        
    \end{minipage}
    \caption{Example of HepScript's grammar (continued).}
\end{figure*}

\FloatBarrier
 \section{Storing Variables for Kinematic Fit}\label{sec:appendix_B}
% Store kmfrit variables
\begin{figure*}[htbp]
    \centering
    \begin{minipage}[t]{0.99\linewidth}
    \includegraphics[width=1.0\linewidth]{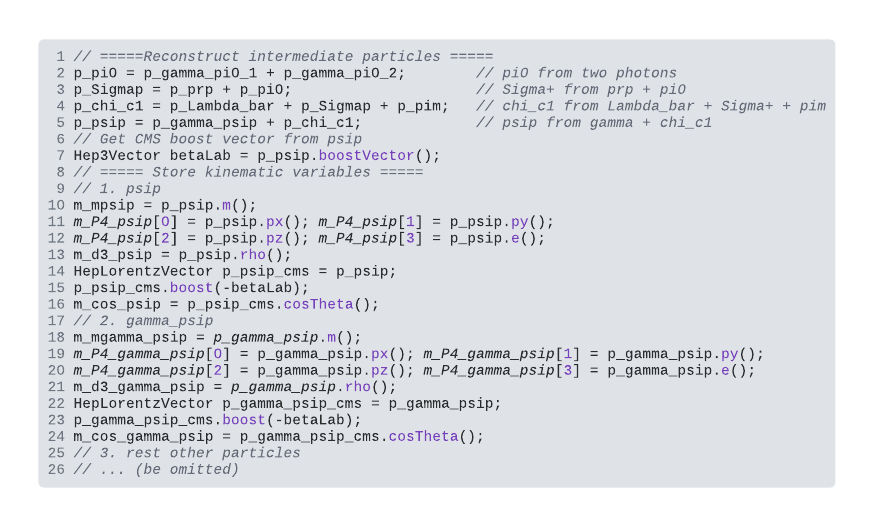}        
    \end{minipage}
    \caption{LLM-generated BOSS code snippets for storing kinematic variables of particles.}
    \label{fig:store_variable}
\end{figure*}

\FloatBarrier
 \section{HepScript Syntax Reference in YARD}\label{sec:appendix_C}
% YARD document example
\begin{figure*}[htbp]
    \centering
    \begin{minipage}[t]{0.99\linewidth}
    \includegraphics[width=1.0\linewidth]{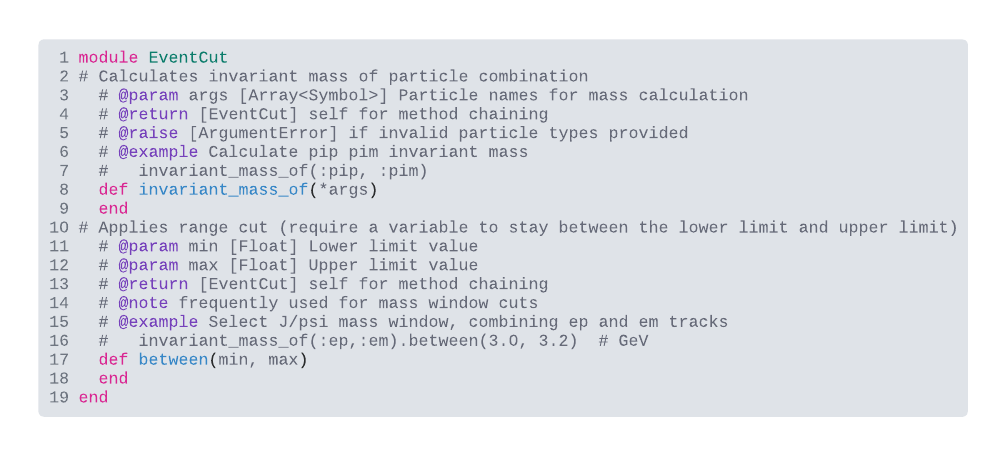}        
    \end{minipage}
    \caption{YARD format documentation for syntax reference. The presented method is for applying invariant mass range requirements.}
    \label{fig:yard_example}
\end{figure*}

\FloatBarrier
 \section{Prompt for Generating HepScript Specification}\label{sec:appendix_D}

%\inputminted{markdown}{code/paper2DSL.md}
\begin{lstlisting}[language={}, basicstyle=\ttfamily\small, breaklines=true, frame=single]
# Role: You are an expert in high energy physics (HEP) data analysis and software engineering, specialized at BESIII experiment and tasked with translating analysis workflows from BESIII published papers into structured Ruby DSL codes."

# Task: Translate HEP Analysis Workflow to Ruby DSL

You are given:
- A High Energy Physics (HEP) paper describing an analysis at BESIII experiment.
- A Ruby DSL (Domain-Specific Language) for expressing dataset preparation and event selection.
- A comprehensive DSL example and a YARD-formatted DSL manual.

Your objective is to translate **only the dataset preparation and event selection parts** of the analysis workflow from the paper into Ruby DSL code, using **only existing DSL capabilities**. Follow the rules and guidelines below strictly.

---

## 1. Scope & Endpoint

### Dataset Preparation
- Write **decay cards** for each signal process as described in the paper. Also Write **decay card** for background process, if the paper mentions or implies that this background is simulated in the analysis. These cards are used to generate exclusive Monte Carlo (MC) samples.
- If a particle is reconstructed via multiple decay modes, create a separate decay card for each mode.
- **Decay card format:** Follow EvtGen syntax and EvtGen particle names. Use `PHSP` generator if the specific generator is uncertain.
- **Top mother particle:** If the paper lacks the a top mother particle (e.g., electron and positron directly produces the signal final states without intermediate mesons like :Jpsi or :psi(2S), in some cross section measurements), use `psi(4260)` as the top mother particle (BESIII convention for simulation) in the decay card.
- **Particle aliasing:** When a particle appears in multiple decay modes within the same card, define an alias to separate them (see example below).
```ruby
   example_decay_card_with_alias = <<~DECAYCARD
      Alias another_phi phi         

      Decay chi_c1
      1.000 phi another_phi HELAMP 0 0 1 0 1 0 0 0 1 0 1 0 0 0;
      Enddecay

      Decay phi
      1.000 K+ K- VSS;
      Enddecay

      Decay another_phi
      1.000 pi+ pi- pi0 PHSP;
      Enddecay

      End
   DECAYCARD
``` 

### Event Selection
- Include **all event selection steps occurring BEFORE or DURING the kinematic fit**.
- After the kinematic fit ends, the energy and momentum of participating particles are automatically corrected. **Exclude any cuts applied to quantities derived from kinematic-fit-corrected variables** (directly or indirectly).
- If multiple kinematic fits exist, stop at the **final fit**. Any cuts after that are out of scope.
- **Kalman kinematic fit:** Do **not** end at a Kalman kinematic fit. Use it only when the paper explicitly mentions a fit to reconstruct intermediate particles like pi0, eta, etc. (Kalman fit typically constrains to a nominal mass of a light meson, which is called "1-C" representing "one constraint on mass". Contrastively, the kinematic fit constrains the four-momentum of the final state system to the center-of-mass energy, which is called "4-C" representing "four constraints on energy-momentum". The DSL provides a method to perform mass constraint on the final state system.)

---

## 2. Translation Guidelines

- Map paper procedures **1:1** to DSL code whenever possible.
- Maintain the original physics meaning for all expressed steps.
- Add concise comments explaining the DSL code.

---

## 3. Translation Process

1. Extract analysis steps from the paper in sequence.
2. For each step, attempt to express it using **current DSL capabilities**.
3. **If the current DSL cannot express a specific procedure or selection criterion, skip that step entirely** (do not comment on the omission).
4. Continue with subsequent steps that **can** be expressed.

---

## 4. Multiple Signal Processes

- If the paper defines multiple signal processes, define separate `Algorithm` instances when needed (such that these process have different final states and selection criteria).
- Use existing DSL patterns for handling multiple datasets.

---

### Special Case: Shared Final States
- If multiple processes share identical final states and selection criteria (e.g., chi_c0, chi_c1, chi_c2), you may create a **single** `Algorithm` instance for all.  The resulting algorithm will apply to all processes because they share final states and selection.
- `with_decay_card()` method is used to generate header files (.h) for an algorithm, defining kinematic variables for particles from the input decay card. One `Algorithm` instance always requires one decay card.

---

## 5. Special DSL Rules (Error Prone)
- Track/photon selection & PID: Functions starting with n (e.g., npip) take a logical expression as a parameter (e.g., ">=2").
- Particle Name Convention Outside Decay Cards: Decay cards use standard EvtGen names (e.g., e+), but Ruby variable names cannot contain special characters like + or -. Therefore, use the mapped symbols below for particle names referenced outside the decay card (e.g., in selection code):

| EvtGen Name    |Ruby DSL Symbol|
|----------------|------------|
| e+             | :ep        |
| em             | :em        |
| mu+            | :mup       |
| mu-            | :mum       |
| pi+            | :pip       |
| pi-            | :pim       |
| K+             | :kp        |
| K-             | :km        |
| p+             | :prp       |
| anti-p-        | :prm       |
| Lambda0        | :Lambda    |
| anti-Lambda0   | :Lambda_bar|
| K_S0           | :K_S0       |
| K_L0           | :K_L0       |
| gamma          | :gamma     |
| pi0            | :pi0       |
| eta            | :eta       |
| anti-n-        | :n_bar     |
| eta'           | :etap      |
| K*0            | :k_star    |
| K*+            | :k_starp   |

- For particles not listed, infer a legal Ruby symbol (e.g., replace special characters with underscores or descriptive names).

---

## 6. Particle List and Candidate Tracking

 During execution of the DSL, each particle type has a corresponding list that stores the indices of reconstructed candidates. If you reference a particle whose list has not yet been created, an error will occur. The following methods create or modify these lists. You must ensure that operations on a particle type only happen after its list has been properly initialized.

### List Creation and Modification Methods

  - Calling `select_track()` method: creating charged track lists, including positively charged tracks (named `iChrgp`) and negatively charged tracks (named `iChrgn`). The indices of selected tracks are stored in these lists.
  - Calling `select_photon()` method: Creating a photon list containing all good photons in the event. Indices are stored in a list typically named `nGood`.
  - Calling `pid()/identify()` method: Creating typed track lists for specific particle hypotheses. There are three possible track types at BESIII experiment that can be distinguished using `identify()`: pion (`:pip`, `:pim`), kaon (`:kp`, `:km`), proton (`:prp`, `:prm`).
     Example" `identify :pip "from_kaon"` creates a list named `index_pip` storing indices of tracks identified as `:pip`.
  - Calling `pid(has_lepton = true)` method: Creating a list of leptons (electrons and muons) in the event. **electron and muon cannot be distinguished using `identify()`**. Instead, you can simply use `pid(has_lepton = true)` to automatically identify leptons. You can also use `pid(has_lepton = true)/identify()` to identify leptons along with pion, kaon, and proton at the same time.
  - Calling `assign()` method: Changing the type of a track list by copying indices from one list to another.
    Example: `assign({:chrgp => :pip, :chrgn => :pim})` copies indices from `iChrgp` to `index_pip` and from `iChrgn` to `index_pim`. If the target lists do not exist, they are created; if they already exist, duplicates are ignored.
  - Calling `remove()` method: Opposite of assign(). It removes particles from the list.
   Example, `remove  ([:prp <= :chrgp,:prm <= :chrgn])` removes proton indices from `iChrgp` and anti-proton indices from `iChrgn`.
  - Calling `remove_used_particle_from_candidate_list()` method: Used only inside a `secondary_vertex_fit` block. It removes indices of particles that have been used in the fit from their respective typed lists, ensuring they are not reused
  - Intermediate particle reconstruction methods: Methods such as `build_virtual_particle()` and `kalman_kinematic_fit` (with mass constraint) create new lists for the reconstructed intermediate particles (e.g., pi0, eta). These lists are populated with the indices of successfully reconstructed candidates.

### Important Note
   - The special lists that combines electron and muon in total are  `index_lp` and `index_lm`, which can be created **only** by calling `pid(has_lepton = true)`. 
   - Always track which particle lists are available at each stage of your translation. If a step in the paper requires using a particle whose list has not yet been created (e.g., because the reconstruction of that particle occurs later), you must either restructure the order (if allowed) or skip that step as per the translation rules. The DSL follows a logical flow: first select tracks/photons, then perform particle identification, then build intermediate particles, and finally apply kinematic fits.

---

## 7. Output Format
- Provide only the DSL code translation in a **single** Ruby code block.
- Do **not** include any explanatory text outside the code block.

---

## 8. Validation Checks (Before Output)
- No post-kinematic fit cuts are included.
- Physics meaning is maintained for expressed steps.
- No DSL extensions are proposed or implemented.
- Skipped steps are omitted entirely (no comments).
- Only existing DSL patterns are used.
- Particle names outside decay cards follow the convention above.
- Particle list dependencies are respected (i.e., no use of a particle before its list is created).

---

# DSL Code Structure Example
Below is the comprehensive DSL example:

{{example}}

# DSL Manual
This is the manual of the DSL in YARD format, please read the event selection part carefully, learn the DSL grammar, and handle its usage before generating the DSL code. Content of the DSL manual:

{{DSL_manual}}

# Paper to be Translated
Now, read the following paper content and generate the corresponding DSL code. When you're uncertain, refer to the DSL manual and the DSL example.
Content of the paper:

{{paper_content}}
\end{lstlisting}

%\FloatBarrier
%\section{Results of Evaluating ROOT Code %Generation}\label{sec:appendix_E}

\end{document}

%% file: def-com.tex
%%************************************************************
%% Shared ones
%%************************************************************
%\newcommand{\BR}{\mathcal{B}}
\newcommand{\afs}{\alpha_s}
\newcommand{\bgp}{\beta\gamma}
\newcommand{\eff}{\varepsilon}
\newcommand{\sintht}{\sin{\theta}}
\newcommand{\costht}{\cos{\theta}}
\newcommand{\dedx}{dE/dx}

%%%%%%%%%%%%%%%%%%%%%%%%%%%%%%%%%%%%%%%%%%event selection
\newcommand{\probfc}{Prob_{\chi^2}}
\newcommand{\probpi}{Prob_{\pi}}
\newcommand{\probka}{Prob_{K}}
\newcommand{\probpr}{Prob_{p}}
\newcommand{\proball}{Prob_{all}}

%%%%%%%%%%%%%%%%%%%%%%%%%%%%%%%%%%%%%%%%%%charmonium
\newcommand{\chicJ}{\chi_{cJ}}
\newcommand{\gchicJ}{\gamma\chi_{cJ}}
\newcommand{\gchica}{\gamma\chi_{c0}}
\newcommand{\gchicb}{\gamma\chi_{c1}}
\newcommand{\gchicc}{\gamma\chi_{c2}}
\newcommand{\hc}{h_c(^1p_1)}
\newcommand{\qqb}{q\bar{q}}
\newcommand{\uub}{u\bar{u}}
\newcommand{\ddb}{d\bar{d}}
\newcommand{\SSB}{\Sigma^+\bar{\Sigma}^-}
\newcommand{\ccb}{c\bar{c}}

%%************************************************************
%% Variables: decay modes of psiprime
%%************************************************************
\newcommand{\psipto}{\psi^{\prime}\rightarrow \pi^+\pi^- J/\psi}
\newcommand{\ptomm}{J/\psi\rightarrow \mu^+\mu^-}
\newcommand{\ppp}{\pi^+\pi^- \pi^0}
\newcommand{\pip}{\pi^+}
\newcommand{\pim}{\pi^-}
\newcommand{\kap}{K^+}
\newcommand{\kam}{K^-}
\newcommand{\ks}{K^0_s}
\newcommand{\pbar}{\bar{p}}
\newcommand{\jp}{J/\psi\rightarrow \gamma\pi^0}
\newcommand{\je}{J/\psi\rightarrow \gamma\eta}
\newcommand{\jep}{J/\psi\rightarrow \gamma\eta^{\prime}}

%%%%%%%%%%%%%%%%%%%%%%%%%%%%%%%%%%%%%%% 2prongs
\newcommand{\LL}{\ell^+\ell^-}
\newcommand{\EE}{e^+e^-}
\newcommand{\MM}{\mu^+\mu^-}
\newcommand{\GG}{\gamma\gamma}
\newcommand{\TT}{\tau^+\tau^-}
\newcommand{\pp}{\pi^+\pi^-}
\newcommand{\kk}{K^+K^-}
\newcommand{\ppb}{p\bar{p}}
\newcommand{\gpp}{\gamma \pi^+\pi^-}
\newcommand{\gkk}{\gamma K^+K^-}
\newcommand{\gppb}{\gamma p\bar{p}}
\newcommand{\ggee}{\gamma\gamma e^+e^-}
\newcommand{\gguu}{\gamma\gamma\mu^+\mu^-}
\newcommand{\ggll}{\gamma\gamma l^+l^-}
\newcommand{\ppee}{\pi^+\pi^- e^+e^-}
\newcommand{\ppuu}{\pi^+\pi^-\mu^+\mu^-}
\newcommand{\etap}{\eta^{\prime}}
\newcommand{\gpi}{\gamma\pi^0}
\newcommand{\geta}{\gamma\eta}
\newcommand{\getap}{\gamma\etap}
%%%%%%%%%%%%%%%%%%%%%%%%%%%%%%%%%%%%%%% 4prongs
\newcommand{\pppp}{\pi^+\pi^-\pi^+\pi^-}
\newcommand{\ppkk}{\pi^+\pi^-K^+K^-}
\newcommand{\pppr}{\pi^+\pi^-p\bar{p}}
\newcommand{\kkkk}{K^+K^-K^+K^-}
\newcommand{\kskp}{K^0_s K^+ \pi^- + c.c.}
\newcommand{\ppkp}{\pi^+\pi^-K^+ \pi^- + c.c.}
\newcommand{\ksks}{K^0_s K^0_s}
\newcommand{\dphi}{\phi\phi}
\newcommand{\phikk}{\phi K^+K^-}
\newcommand{\ppeta}{\pi^+\pi^-\eta}
\newcommand{\gpppp}{\gamma \pi^+\pi^-\pi^+\pi^-}
\newcommand{\gppkk}{\gamma \pi^+\pi^-K^+K^-}
\newcommand{\gpppr}{\gamma \pi^+\pi^-p\bar{p}}
\newcommand{\gkkkk}{\gamma K^+K^-K^+K^-}
\newcommand{\gkskp}{\gamma K^0_s K^+ \pi^- + c.c.}
\newcommand{\gppkp}{\gamma \pi^+\pi^-K^+ \pi^- + c.c.}
\newcommand{\gksks}{\gamma K^0_s K^0_s}
\newcommand{\gphiphi}{\gamma \phi\phi}

%%%%%%%%%%%%%%%%%%%%%%%%%%%%%%%%%%%%%%% 6prongs
\newcommand{\tpp}{3(\pi^+\pi^-)}
\newcommand{\tppkk}{2(\pi^+\pi^-)(K^+K^-)}
\newcommand{\pptkk}{(\pi^+\pi^-)2(K^+K^-)}
\newcommand{\tkk}{3(K^+K^-)}
\newcommand{\gtpp}{\gamma 3(\pi^+\pi^-)}
\newcommand{\gtppkk}{\gamma 2(\pi^+\pi^-)(K^+K^-)}
\newcommand{\gpptkk}{\gamma (\pi^+\pi^-)2(K^+K^-)}
\newcommand{\gtkk}{\gamma 3(K^+K^-)}

%%************************************************************
%%  Variables and Formula
%%************************************************************
\newcommand{\psp}{\psi(3686)}
\newcommand{\jpsi}{J/\psi}
\newcommand{\ar}{\rightarrow}
\newcommand{\lra}{\longrightarrow}
\newcommand{\jpsito}{J/\psi \rightarrow }
\newcommand{\ptoppjp}{J/\psi \rightarrow\pi^+\pi^- J/\psi}
\newcommand{\pspto}{\psi^\prime \rightarrow }
\newcommand{\ptop}{\psi'\rightarrow\pi^0 J/\psi}
\newcommand{\ptoeta}{\psi'\rightarrow\eta J/\psi}
\newcommand{\ecto}{\eta_c \rightarrow }
\newcommand{\ecpto}{\eta_c^\prime \rightarrow }
\newcommand{\xto}{X(3594) \rightarrow }
\newcommand{\chicJto}{\chi_{cJ} \rightarrow }
\newcommand{\chiczto}{\chi_{c0} \rightarrow }
\newcommand{\chicoto}{\chi_{c1} \rightarrow }
\newcommand{\chictto}{\chi_{c2} \rightarrow }
\newcommand{\pspp}{\psi^{\prime\prime}}
\newcommand{\ptochic}{\psi(2S)\ar \gamma\chi_{c1,2}}
\newcommand{\ppjpsi}{\pi^0\pi^0 J/\psi}
\newcommand{\utoeta}{\Upsilon^{\prime}\ar\eta\Upsilon}
\newcommand{\ww}{\omega\omega}
\newcommand{\wf}{\omega\phi}
\newcommand{\ff}{\phi\phi}
\newcommand{\npsp}{N_{\psp}}
\newcommand{\llb}{\Lambda\bar{\Lambda}}
\newcommand{\llbpi}{\llb\pi^0}
\newcommand{\llbeta}{\llb\eta}
\newcommand{\ppi}{p\pi^-}
\newcommand{\pbpi}{\bar{p}\pi^+}
\newcommand{\lamb}{\bar{\Lambda}}
%%=================================================
%% abbreviated commands of LaTeX
%%=================================================
\def\ctup#1{$^{\cite{#1}}$}
\newcommand{\bfg}{\begin{figure}}
\newcommand{\efg}{\end{figure}}
\newcommand{\bitm}{\begin{itemize}}
\newcommand{\eitm}{\end{itemize}}
\newcommand{\bnum}{\begin{enumerate}}
\newcommand{\enum}{\end{enumerate}}
\newcommand{\btbl}{\begin{table}}
\newcommand{\etbl}{\end{table}}
\newcommand{\btbu}{\begin{tabular}}
\newcommand{\etbu}{\end{tabular}}
\newcommand{\bcl}{\begin{center}}
\newcommand{\ecl}{\end{center}}
\newcommand{\bbt}{\bibitem}
\newcommand{\beq}{\begin{equation}}
\newcommand{\eeq}{\end{equation}}
\newcommand{\beqr}{\begin{eqnarray}}
\newcommand{\eeqr}{\end{eqnarray}}
%%===========================================
%%  color setting
%%===========================================
\newcommand{\red}{\color{red}}
\newcommand{\blue}{\color{blue}}
\newcommand{\yellow}{\color{yellow}}
\newcommand{\green}{\color{green}}
\newcommand{\purple}{\color{purple}}
\newcommand{\brown}{\color{brown}}
\newcommand{\black}{\color{black}}

%%===========================================
%%  For revised
%%===========================================
\definecolor{boslv}{rgb}{0.0, 0.65, 0.58}%persiangreen
\definecolor{Munsell}{HTML}{00A877}
\newcommand{\psip}{\psi^{'}}
\newcommand{\psipp}{\psi(3686)}

\newcommand{\Br}{\mathcal{B}}
\newcommand{\too}{\rightarrow}
\newcommand{\del}{\color{red}\sout}
\newcommand{\new}{\color{blue}\uwave}

%% file: Acknowledgement.tex
\begin{ack}
This project is supported by the Strategic Priority Research Program of Chinese Academy of Sciences under Grant XDA0480600; Program of Science and Technology Development Plan of Jilin Province of China under Contract No. 20230101021JC. The authors declare no competing interests. We would like to thank the BESIII Collaboration for their support on test datasets and softwares; the IHEP computing center for their support on computing resources; Yu-Zhe Shi (from School of Advanced Manufacturing and Robotics, PKU), Kun He and Ming-Chen Liu (from School of Computer Science and Technology, HUST) for their innovative, pioneering works on automatical DSL design, which inspires us to develop HepScript; Jie Liu, Liang-Yi Kang, and Shuai Wang (from Institute of Software, CAS) for the helpful discussions.
\end{ack}